\documentclass{article}

\usepackage{PRIMEarxiv}
\usepackage[utf8]{inputenc} 
\usepackage[T1]{fontenc}    
\usepackage{hyperref}       
\usepackage{url}            
\usepackage{booktabs}       
\usepackage{amsfonts}       
\usepackage{nicefrac}       
\usepackage{microtype}      
\usepackage{lipsum}
\usepackage{fancyhdr}       
\usepackage{graphicx}       
\graphicspath{{media/}}     
\usepackage{amsmath} 
\usepackage{subcaption}
\usepackage{multirow}
\usepackage{hhline}
\usepackage{makecell, booktabs, caption}
\usepackage{subcaption}
\usepackage{tabularray}
\usepackage{colortbl}
\definecolor{light-gray}{gray}{0.95}
\usepackage{colortbl} 
\usepackage{geometry} 
\usepackage[table]{xcolor} 
\usepackage{adjustbox} 
\renewcommand{\tabcolsep}{1pt}
\ifdefined\Vcolumn \else
    \newcolumntype{B}{!{\vline width 0.150pt}}
    \newcolumntype{C}[1]{>{\centering\arraybackslash}m{#1}}
    \newcommand{\Vcolumn}{}
\fi

\title{Towards Deep Learning for Predicting Microbial Fuel Cell Energy Output
}

\author{
  Adam Hess-Dunlop \\
  Affiliation \\
  Arizona State University \\
  \texttt{ahessdun@asu.edu} \\
   \And
  Harshitha Kakani \\
  Affiliation \\
  UC Santa Cruz \\
  \texttt{ckakani@ucsc.edu} \\
   \And
  Colleen Josephson \\
  Affiliation \\
  UC Santa Cruz \\
  \texttt{cjosephson@ucsc.edu} \\
}

\begin{document}
\maketitle

\begin{abstract}
Soil microbial fuel cells (SMFCs) are an emerging technology which offer clean and renewable energy in environments where more traditional power sources, such as chemical batteries or solar, are not suitable. With further development, SMFCs show great promise for use in robust and affordable outdoor sensor networks, particularly for farmers. One of the greatest challenges in the development of this technology is understanding and predicting the fluctuations of SMFC energy generation, as the electro-generative process is not yet fully understood. Very little work currently exists attempting to model and predict the relationship between soil conditions and SMFC energy generation, and we are the first to use machine learning to do so. In this paper, we train Long Short Term Memory (LSTM) models to predict the future energy generation of SMFCs across timescales ranging from 3 minutes to 1 hour, with results ranging from 2.33\% to 5.71\% MAPE for median voltage prediction. For each timescale, we use quantile regression to obtain point estimates and to establish bounds on the uncertainty of these estimates. When comparing the median predicted vs. actual values for the total energy generated during the testing period, the magnitude of prediction errors ranged from 2.29\% to 16.05\%. To demonstrate the real-world utility of this research, we also simulate how the models could be used in an automated environment where SMFC-powered devices shut down and activate intermittently to preserve charge, with promising initial results. Our deep learning-based prediction and simulation framework would allow a fully automated SMFC-powered device to achieve a median 100+\% increase in successful operations, compared to a naive model that schedules operations based on the average voltage generated in the past.
\end{abstract}


\begin{figure}[ht]
    \centering
    \includegraphics[width=0.52\textwidth]{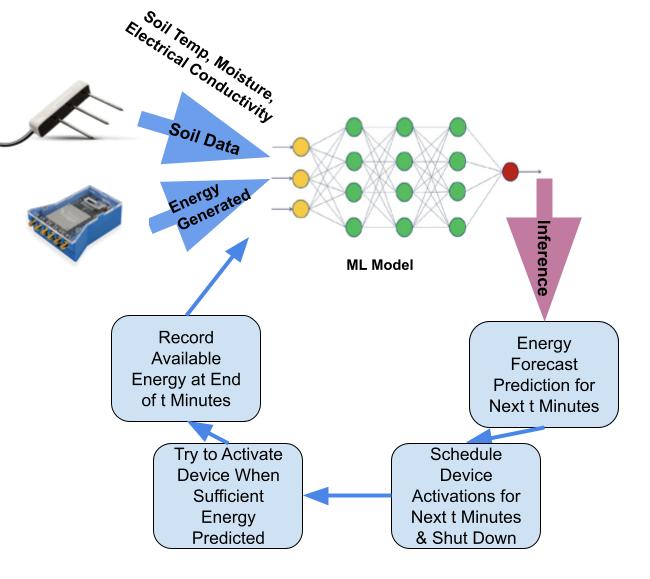}
\caption{Overview and Vision. This work aims to create a predictive model for scheduling the activation of SMFC powered devices, allowing them to activate intermittently and then shut down to conserve energy.} 
    \label{fig:vision}
\end{figure}

\section{Introduction}
Climate change is already affecting every aspect of our society, widening existing socioeconomic disparities across the world. Some of the most dangerous changes are occurring in our global food systems, where extreme weather has made feeding our growing population a challenge. Data-driven agriculture techniques, such as moisture and nutrient monitoring, have enabled us to grow more food while using fewer resources \cite{ZOTARELLI200923}. Unfortunately, the adoption rates for the sensor networks that enable data-driven agriculture remain low \cite{RAJAK2023100776} investigated. Part of the reason for this is that deploying and maintaining sensor networks is currently costly and labor intensive, as farms and other managed lands typically lack robust power and communication infrastructure. One promising area of investigation involves harvesting power for sensor networks from the soil itself \cite{10.1145/3568113.3568117}. Soil microbial fuel cells (SMFCs) are a compact, low-cost bioelectrochemical system that harvest power from \emph{exoelectrogenic microbes} that occur naturally in the soil.

Though SMFCs only produce microwatts of power, electronics have progressed such that this is actually enough to power the latest generation of ultra-low power devices\cite{10.1145/3460112.3472326, 10.1145/3631410, 10.1145/3560905.3568525}. SMFCs are a renewable source of energy, and unlike more traditional power sources (chemical battery, solar, etc.), they are typically constructed using environmentally inert materials with a very low carbon footprint. Once fully developed, they could power outdoor sensor networks, giving farmers access to high-resolution, real-time data on their fields towards making educated decisions on farm management. They also have the potential to be biosensors in and of themselves\cite{OLIAS2020137108}. For example,  the electricity generated by the microbial communities can be used as a signal to indicate heavy-metal contamination or dissolved oxygen in water\cite{ABBAS2022135036, bios13010145}.

A key challenge with leveraging this unique source of biopower is that the electro-generative process is not yet fully understood, with rises and drops in energy production being common due to a variety of complex factors, including temperature, soil type, moisture and more. This makes SMFCs difficult to use as a source of consistent and reliable power. To address this barrier, we have created a deep learning model to predict SMFC energy generation over time, increasing their viability as an energy source for low-power applications. To our knowledge this is the first work to predict SMFC energy generation using deep learning.

In addition to making point estimates of future energy output, our work models the uncertainty of these estimates using quantile regression, as defined in Section ~\ref{quant}. This allows us to make conservative estimates of future energy generation when necessary, in order to minimize the possibility of a device not having enough energy to perform the operations our model predicts it will. It is often necessary for low-power applications, such as outdoor sensor networks, to shut down for periods of time to gather energy, activating only when necessary to perform operations \cite{10.1145/3485730.3493363}. This is known as \emph{intermittent computing} \cite{lucia_et_al:LIPIcs.SNAPL.2017.8}. Our approach is key for the types of intermittently active, low-power applications supported by SMFCs, where every microwatt makes a difference, and trying to activate a device before enough energy is stored would waste precious energy. This approach could also be useful for intermittently active, low-power applications supported by small solar cells, as well as for use in power grids to plan for "worst case scenarios" where little energy is available.
Our model, which has been trained and evaluated on months of real SMFC data, 
predicts performance for future time horizons using recently observed data. These predictions make it possible to schedule device operations ahead of time, allowing for more effective resource allocation. For example, 
it would be possible to adjust the duty cycle of wireless data transmission---an operation with high power consumption---based on the predicted power budget. In times of low energy availability, for example, the number of attempted wireless data transmissions could be reduced to conserve energy for more essential operations (e.g. timekeeping, local data logging). 

There are multiple advantage of predictive models over a naive approach that uses a fixed duty cycle. The first advantage is that overall system downtime can be reduced. When no prediction information is available, the only option to maximize longevity is reducing the frequency of operation to as low as acceptably possible. With a predictive model, the system can to take advantage of times of high energy availability to perform more frequent and/or sophisticated operations (e.g. over-the-air firmware updates). The second advantage is that we can avoid wasting energy. Intermittent computing applications should only activate only when there is enough energy available to perform the desired operations, e.g. transmitting a packet. If the operation is not successfully completed, then no useful progress is made, and the stored energy is wasted and unavailable to use in a future potentially-successful operation. Our model addresses this need by allowing for a lower-bound prediction of energy generation using quantile regression, as defined in Section ~\ref{quant}. Likewise, we introduce three unique metrics in this paper to evaluate the usefulness of our model for intermittent computing applications. This is in addition to the more standard Mean Average Percent Error (MAPE), which is used to measure the overall accuracy of the model’s predictions compared to the ground truth values. The domain-specific metrics in this paper include the Failed Prediction and Overestimation rates, designed to measure how often the model predicts greater energy generation than the true value, and the Missed Activation rate, designed to predict how many more times a device could have been activated by using a theoretical"perfect model” capable of predicting exact energy generation. These metrics are defined in further detail in Section~\ref{sec:eval}

\section{Background}
\label{sec:headings}

\subsection{Microbial Fuel Cells}
\begin{figure}[h]
    \centering
    \includegraphics[width=0.5\textwidth]{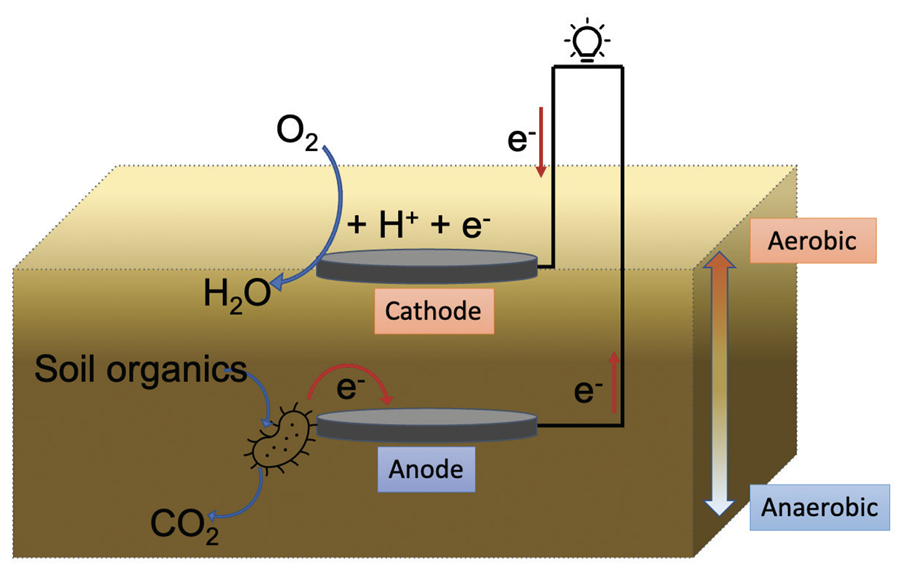}
\caption{Visual diagram of soil microbial fuel cell electrogenerative process.}
\label{fig:electro_gen}
\end{figure}

Most generally, microbial fuel cells (MFCs) are electrochemical cells that generate electricity from the transfer of electrons resulting from microbial interactions. Soil microbial fuel cells (SMFCs) use the microbial  interactions within soil, but other types of MFCs can use wastewater or sediment as well \verb+\citet+{10.1145/3568113.3568117}. Two key requirements for SMFCs to operate are anaerobic conditions and a sufficient presence of soil organic matter. Certain types of microbes, known as exoelectrogens, produce a spare electron as part of their natural respiration process. By placing an anode within the soil connected to a cathode outside the soil, the anode can receive these electrons from the soil microorganisms and allowing them to flow to the cathode, generating electricity. This process is described visually in Figure ~\ref{fig:electro_gen}, adapted from Josephson et al. \cite{10.1145/3568113.3568117}.

\subsection{Quantile Regression in Energy Forecasting}
Compared to other types of models, deep learning can be used to make extremely accurate predictions in various application areas. However, the non-statistical nature of deep learning makes model interpretation difficult. Quantile regression is one way to address this weakness by allowing deep learning models to obtain prediction intervals as well as point estimates. Previous works have used machine learning and deep learning with quantile regression to forecast renewable energy generation \cite{WANG201910}\cite{XU2023107772}.  However, our work is the first to use this technique to predict the energy output of SMFCs. Forecasting energy output for SMFCs is more difficult than for more traditional sources (e.g. wind, solar) for several reasons. SMFC energy generation is directly tied to microbial conditions, which are governed by complex biological processes that are not yet fully understood. For example, it can take microbial communities longer to respond to changes in environmental conditions than solar or wind systems, resulting in delayed and unpredictable changes to energy generation \cite{10.1145/3568113.3568117}.

\subsection{SMFC Modeling for Incubation Phase}
Before being deployed in the field, SMFCs usually undergo an incubation phase within an indoor environment. Dziegielowski et al. \cite{DZIEGIELOWSKI2023103071} created a physics-based mathematical model to predict the voltage of SMFCs during their incubation phase based on initial soil conditions. However, this work only predicts voltage (not current).

Averaging across all three soil types used in the model, 82.7$\%$ of predictions had a relative error (defined as the absolute error divided by the experimental value) of less than 10$\%$, and 71.7$\%$ of predictions had a relative error of less than 5$\%$ \cite{DZIEGIELOWSKI2023103071}. However, there are several key differences between the physics-based model developed by Dziegielowski et al. and the one presented in our paper, which makes direct comparison of the models difficult. The physics-based model was validated on the same data used for training, while the data for our model is split into training, validation, and testing sets, used to fit the model, tune hyper-parameters, and evaluate performance, respectively. Furthermore, the physics-based model directly simulates the biophysical conditions of an SMFC over the course of several months in order to predict the output, starting from the initial conditions of the soil. In contrast, our model accounts for the rapidly-changing conditions of a non-laboratory deployment by periodically reading in sensor data to update the soil conditions, and uses this to predict SMFC at various time horizons into the future. 

Overall, \cite{DZIEGIELOWSKI2023103071} offers strong insight into soil conditions that promote high voltage output for SMFCs during incubation, but does not attempt to model these relationships in an out-of-lab deployment setting. 
At the time of this writing, our work is the first to (1) model and predict SMFC performance during field deployment outside of the lab, (2) predict current as well as voltage, and (3) predict for multiple time horizons with bounds on the uncertainty.  

Previous works have  \cite{DZIEGIELOWSKI2023103071}. However, this work only predicts voltage (not current), and the time horizon of the prediction is unclear. The metrics used to evaluate ~\cite{DZIEGIELOWSKI2023103071}'s model are not directly comparable to ours, and their source code is not available, but they generally appear to have "relative error" of 5-10\%, vs. 2.33-5.71\% MAPE for our model's voltage predictions. At the time of this writing, our work is the first to (1) model and predict SMFC performance during field deployment outside of the lab, (2) predict current as well as voltage, and (3) predict for multiple time horizons with bounds on the uncertainty.  

\subsection{Intermittently Active, SMFC-Powered Devices}

Marcano et al. \cite{10.1145/3485730.3493363} successfully developed a low-power, e-ink display device powered exclusively by an SMFC, demonstrating the potential of this technology as a renewable energy source. This device was designed to be active intermittently: it is turned off while charging, and manually powered on once enough energy had been stored for activation. However, the evaluation was limited to laboratory settings where unrealistically high moisture levels were necessary to sustain the device. 

Yen et al. \cite{10.1145/3631410} created a proof-of-concept system that successfully powers an RF backscatter tag. They also developed a framework for calculating the number of operations various SMFC-powered devices can perform based on measured SMFC voltage traces, accounting for the complex harvesting process required to use SMFC-generated energy. 

Our work focuses on modeling and predicting MFC energy output. This allows us to develop a framework for scheduling intermittent computing device operations in advance, so devices can activate only when they have enough energy to do so, thereby conserving energy. This would enhance the functionality of the prototypes discussed in \cite{10.1145/3631410} and \cite{10.1145/3485730.3493363}, the latter of which required a human participant to monitor energy availability and manually activate the device.

\subsection{Task Scheduling Frameworks for Intermittently Powered Systems}
This paper is the first work, to our knowledge, to present a deep learning based approach to predicting SMFC power output, and scheduling useful tasks and operations based on these predictions. However, other works have created task scheduling frameworks for intermittently powered systems not fueled by SMFCs. Zygarde is one such framework, which has been validated on four standard datasets---three consisting of labeled images for training machine learning models, and one containing audio files---and deployed in six real-world application settings, with RF and solar-powered systems \cite{Zygarde}. The applications used for validation are performed on audio and visual tasks, so we are unable to compare their performance to our regression-based model. 

Furthermore, Zygarde does not directly predict energy generation, but rather uses a uses a probabilistic method to model energy randomness. In contrast, our work uses deep neural networks for regression to directly predict voltage, current and power output, as well as the upper and lower bounds for these values.

\section{Methodology and Design}
\subsection{Dataset Description} \label{dataset}
The dataset used to train the models in this paper is taken from a cell that began operation in 2019 and was retired in 2022. Specifically, we use the segment of the data during which the cell was deployed in a field setting, which lasted from June 4th, 2021 to January 6th, 2022. 

We use the first 70\% of the data for our training set, the subsequent 15\% of the data for our validation set, and the last 15\% of the data for our test set. Our input values consist of the average values of various features over a desired time interval, as described in Section ~\ref{preprocess}. Therefore, the size of our sets vary in size from 2559 to 51400 values for the training set and 548 to 11015 values for the validation and testing set, with larger values corresponding to shorter time intervals.
\subsection{Data Pre-Processing and Model Features}~\label{preprocess}
Our models aim to forecast the power generation of an SMFC over five different timescales---3 minutes, 5 minutes, 15 minutes, 30 minutes, and 60 minutes in the future. These timescales were selected to explore how the accuracy of our model changes over different prediction horizons. We begin by sanitizing our data (removing occasional erroneous zero measurements due to data outages), and then resampling it by taking the average of each feature across the timescale for which the current model is being trained. For example, if we want to train a model that predicts the average SMFC power generation over the next hour, we resample our data to obtain the one-hour average for each of the model's features. The features we use to train the model are the power, current and voltage of the SMFC, and the electrical conductivity, temperature, and raw volumetric water content of the soil. The power, current and voltage values are gathered using the open-source RocketLogger system ~\cite{7927164}, and the electrical conductivity, temperature, and raw volumetric water content values are gathered using the commercial TEROS-12 sensor~\cite{Teros-12}.  These values are sampled every 12-15 seconds.

Next we shift the data such that at each timestamp, the model is given access to the features of the previous three time intervals in order to predict the average power, voltage and current for the current time interval. Finally, we add the number of days since SMFC deployment, as well as the hour of the day, to our list of features. The input features are not normalized.

\subsection{Model Building}
This work uses a Long Short-Term Memory (LSTM) model to predict future energy generation for LSTMs. LSTMs are one of the most-used deep learning models for time series data, data which is indexed and processed in temporal order. Many past studies have used LSTMs to predict solar and wind energy generation \cite{YING2023135414} \cite{SRIVASTAVA2018232}, and we apply this approach to SMFC energy forecasting for the first time.

Our model uses the Adam optimizer at the default learning rate, and a sequence length of four for the input data. We change our batch size depending on the desired timescale of the prediction, with a batch size of 300 for 3 minutes, 150 for 5 minutes, 50 for 15 minutes, 20 for 30 minutes, and 8 for 60 minutes. More information on the structure of our model can be found in the open-source Github repository for this project~\cite{MFC_MODEL}.

\subsubsection{Quantile Regression}\label{quant}
While deep learning models excel in making accurate predictions, it is difficult to calculate the uncertainty in the predictions of these models, limiting their usefulness in real-world settings~\cite{10.1007/s10462-023-10562-9}. One method to establish bounds on uncertainty in deep learning models is quantile regression, which allows us to train multiple models, which each explicitly predict a different quantile of the data. For example, if we wish to predict the energy generation at a given time and establish bounds on uncertainty with a 90$\%$ confidence interval, we would train three separate models using quantile regression: one model to obtain point estimates by predicting the median quantile, one model to obtain the lower bound of the prediction interval by predicting the 5th quantile, and another model to obtain the upper bound of the prediction interval by predicting the 95th quantile \cite{WANG201910}. To train these models, we use the following loss function, also known as the "pinball loss" during training:
{\small
\begin{align*}
&\texttt{if predicted} \leq \texttt{actual:}\\
& \hspace{4em} \texttt{loss} = \alpha * \texttt{(actual - predicted)}\\ 
&\texttt{else:} \\
& \hspace{4em} \texttt{loss} = (1 - \alpha) * \texttt{(predicted - actual)}\\ 
\end{align*}
}

\noindent where \texttt{$\alpha$} is the desired quantile, \texttt{predicted} is the predicted current, voltage, and power output for the SMFC, and \texttt{actual}  is the actual current, voltage, and power output for the SMFC. 

Since performing gradient descent requires a differentiable loss function, quantile regression cannot typically be used with deep learning. However, the pinball loss function allows us to overcome this limitation and use quantile regression to quantify the uncertainty of our models ~\cite{rodrigues2018expectation,10.3150/10-BEJ267}. Several other works have also used quantile regression with pinball loss to quantify uncertainty for deep learning-based energy forecasting models ~\cite{WANG201910, XU2023107772}. 

It should be noted that the data does not follow a normal distribution, so the quantiles of our upper and lower bounds will not be equidistant from the median. Furthermore, as the model produces three outputs with different distributions (current, voltage, and power), the outputs are expected to diverge somewhat from the desired quantile, since the model training optimizes the average loss metric of the three.

\subsection{Model Evaluation} \label{sec:eval}
We use our test set, as described in Section ~\ref{dataset}, to evaluate our model. There are several techniques that we can use to interpret the prediction and the accuracy of our models:

\subsubsection{MAPE}
Model accuracy for regression tasks can be calculated using Mean Average Percent Error (MAPE), defined as follows: 
{\small
\begin{equation*}
\frac{1}{n}\sum^{n}\frac{|\texttt{actual} - \texttt{predicted}|}{|\texttt{actual}|}\times 100
\end{equation*}
}

Intuitively, this value is the average percent difference between the predicted values and the actual values in a model. There is no universally agreed upon acceptable value for MAPE, but a MAPE of 0$\%$ would indicate that the model predicts the data with zero error, whereas an MAPE of 50$\%$ indicates that on average, the predicted value is off by 50$\%$ of the actual value.

\subsubsection{Total Energy \% Error} \label{Total_E_E} This metric is the percent difference between the actual and predicted values of the total energy generated. 
Energy is the integral of power over time. The difference between predicted energy income and actual energy income is a more valuable metric than MAPE because the actual magnitude of over-predicting and under-predicting power does not actually matter if the over- and under-predictions compensate for each other. The results for the  Total Energy \% Error metric are recorded in Table ~\ref{tab:model_perf}, with negative values indicating underestimation and positive values indicating overestimation.

\subsubsection{Failed Activation Rate} \label{Fail}
The failed activation rate metric measures how often our model schedules a device activation when there is not enough energy available, resulting in the device failing to activate and the stored energy being wasted. The failed activation rate will be calculated as follows: 
{\small
\begin{equation*}
 \frac{\texttt{failed\_active}}{\texttt{active\_pred}}
\end{equation*}
}

\texttt{failed$\_$active} is the total number of times our model schedules a device activation when there is not enough energy available, resulting in the device failing to activate.
\texttt{active$\_$pred} is the total number of activations scheduled by our model. 

\subsubsection{Missed Activation Rate} \label{Miss}
Unlike overestimation and failed activations, the missed activation rate metric measures how many more times the device could have successfully been activated, if we had access to a theoretical "oracle" model that could perfectly predict and make use of the available energy. It will be calculated as follows: 
{\small
\begin{equation*}
\frac{\texttt{missed\_active}}{\texttt{max\_active}}
\end{equation*}
}

\texttt{missed$\_$active} is the number of additional times the device could been have activated if the model had perfectly predicted how much energy would be available. It is calculated as the theoretical maximum possible number of operations, minus the number of successful operations scheduled by our model.
\texttt{max$\_$active} is the theoretical maximum possible number of operations.

\subsection{Comparison Models} \label{comp_models}
In addition to the rest of the metrics described in this section, we evaluate our scheduling framework by comparing it to two different models: the naive model and the oracle model.

The naive model, also referred to as the naive fixed-duty cycle, is used as a "baseline" model to compare our deep learning-based model to. It operates in the same way as the runtime simulation code described in section ~\ref{runtime_sim}, by using predictions of SMFC voltage to estimate how much usable energy the SMFC will have access to in the future, and then using these estimates to schedule activation of a device. However, while our runtime simulation predicts the voltage using deep learning models, the naive model simply takes the average voltage over the past $x$ days before the start of the test set, with $x$ being the size of the test set.

In contrast, the oracle model is designed to measure the maximum possible number of times we could activate a device using the energy generated during the duration of the test set. This allows us to compare our scheduling framework against a theoretical, perfect maximum. To obtain this maximum number of activations, we simply measure the energy generated in the test set, and divide by the energy required to activate our device.

\subsection{Runtime Simulation} \label{runtime_sim}
Yen et al. \cite{10.1145/3631410} have developed a framework for calculating the number of operations various SMFC-powered devices can perform based on measured SMFC voltage traces, accounting for the complex harvesting process required to use SMFC-generated energy. Our work adapts this simulation to use the voltage predictions from our LSTM models, using publicly available python code created by Yen et al. \cite{10.1145/3631410} \cite{PRACTICAL}. To evaluate our model's performance, we investigate how many times an SMFC could be used to activate the Cinamin beacon, a low-power device designed exclusively to send BLE advertisement packets ~\cite{10.5555/2893711.2893793}. On average, this device requires 3.9 $\mu J$ to activate \cite{PRACTICAL}.  We also make the following modifications to the code presented by Yen et al.: 

\begin{itemize}
    \item The original simulation code uses an estimate of SMFC internal resistance calculated using the one-resistance method used in Fujinaga et al. \cite{IR}. We are unable to use this method for the cell which gathered the data used in this paper,since the deployment ended in 2022 and we no longer have access to the cell. However, the cell we used to gather our data is very similar to the v0 cell used in \cite{10.1145/3631410}, so we use the same internal resistance of 6926$\Omega$ for our calculations.
    \item Furthermore, in the absence of a better estimator, a flat harvester efficiency of 60$\%$ is used for calculations, based on the lowest efficiency found in the ADP5091/ADP5092 energy harvester datasheet for \texttt{V\_in} = 0.5V, \texttt{V\_SYS} = 3V, and a lower range of input voltages of 0.01V~\cite{DATASHEET}.
\end{itemize}

\begin{table*}[ht!]
\renewcommand{\tabcolsep}{1pt}
    \centering
    \caption{\bfseries Model Performances. L, U are upper (95th percentile) and lower bound (5th percentile) estimates for the models; M signifies the median. Total energy error is \% difference between ground truth and predicted values of the total energy generated. The "Failed Activations" metric measures how often our model schedules a device activation when there is not enough energy. The "Missed Activations" metric measures how many additional times the device could have successfully been activated if operating with perfect knowledge. More information on bound estimates and metrics can be found in Secs.~\ref{quant}, ~\ref{Total_E_E}, ~\ref{Fail}, and ~\ref{Miss}.}
    \vspace{1em}
    \begin{adjustbox}{max width=\textwidth}
    \begin{tabular}{|c|c|c|c|c|c|c|c|c|c|c|c|c|c|c|c|}
        \hline
        Possible Activations & \multicolumn{15}{c|}{6023} \\  
        \hline
        Times (Minutes) & \multicolumn{3}{c|}{3} & \multicolumn{3}{c|}{5} & \multicolumn{3}{c|}{15} & \multicolumn{3}{c|}{30} & \multicolumn{3}{c|}{60} \\ 
        \hline 
        Batch Size & \multicolumn{3}{c|}{300} & \multicolumn{3}{c|}{150} & \multicolumn{3}{c|}{50} & \multicolumn{3}{c|}{20} & \multicolumn{3}{c|}{8} \\ 
        \hline 

        Bound Estimate & \cellcolor[HTML]{EFEFEF} & \cellcolor[HTML]{DCDCDC} & \cellcolor[HTML]{C0C0C0} & \cellcolor[HTML]{EFEFEF} & \cellcolor[HTML]{DCDCDC} & \cellcolor[HTML]{C0C0C0} & \cellcolor[HTML]{EFEFEF} & \cellcolor[HTML]{DCDCDC} & \cellcolor[HTML]{C0C0C0} & \cellcolor[HTML]{EFEFEF} & \cellcolor[HTML]{DCDCDC} & \cellcolor[HTML]{C0C0C0} & \cellcolor[HTML]{EFEFEF} & \cellcolor[HTML]{DCDCDC} & \cellcolor[HTML]{C0C0C0} \\
        (Lower, Middle, Upper) & L & M & U & L & M & U & L & M & U & L & M & U & L & M & U \\ 
        \hline     
        Predicted Activations & 6011 & 6014 & 6101 & 6007 & 6014 & 6115 & 6007 & 6043 & 6488 & 6003 & 6123 & 7091 & 6008 & 6193 & 7753 \\[1pt] 
        \hline 
        Failed Activations (\%) & 0.067\% & 2.345\% & 18.866\% & 0\% & 3.242\% & 18.872\% & 0.366\% & 7.744\% & 53.067\% & 0\% & 17.050\% & 97.038\% & 0.566\% & 21.411\% & 99.020\% \\[1pt] 
        \hline 
        Missed Activations (\%) & 0.266\% & 2.490\% & 17.815\% & 0.266\% & 3.894\% & 17.632\% & 0.631\% & 7.438\% & 49.444\% & 0.332\% & 15.673\% & 96.513\% & 0.897\% & 19.343\% & 98.954\% \\[1pt] 
        \hline
        Total Energy Error & -40.674\% & 8.490\% & 24.100\% & -33.987\% & 2.288\% & 42.172\% & -14.651\% & 16.049\% & 27.331\% & -18.220\% & 9.851\% & 28.682\% & -21.772\% & 5.543\% & 35.316\% \\[1pt] 
        \hline
        Test MAPE Power & 38.976\% & 18.748\% & 26.801\% & 34.309\% & 12.486\% & 47.368\% & 16.760\% & 17.121\% & 27.602\% & 20.605\% & 9.678\% & 26.165\% & 23.360\% & 10.964\% & 32.182\% \\ 
        \hline 
        Test MAPE Voltage & 7.291\% & 2.326\% & 9.329\% & 6.422\% & 3.034\% & 9.584\% & 4.771\% & 3.636\% & 12.277\% & 14.407\% & 4.984\% & 14.305\% & 18.970\% & 5.709\% & 17.578\% \\ 
        \hline 
        Test MAPE Current & 37.827\% & 16.210\% & 28.484\% & 31.629\% & 8.906\% & 20.852\% & 18.364\% & 4.267\% & 15.748\% & 14.107\% & 3.069\% & 8.192\% & 14.092\% & 5.016\% & 12.437\% \\ 
        \hline
    \end{tabular}
    \end{adjustbox}
    \label{tab:model_perf}
\end{table*}

\subsection{Time-Series Cross Validation} \label{tscv}
In time-series analysis, one cannot train a model on data gathered after data from the validation or testing sets. Because of this, traditional k-folds cross validation, a method for testing the generalizability of a deep learning model, cannot be used. However, a modified version of k-folds called time-series cross validation can be used to test how well a time-series model generalizes to new data \cite{TSCV}. Using this method, we train and test the performance of our deep learning model with 4 different distributions of training, validation, and test datasets. The first of these uses the first 20\% of the data to train the model, the next 10\% to generate the average voltage used in the naive model, as described in section ~\ref{comp_models}, and the next 10\% as the test set to evaluate model performance. The second dataset distribution uses the first 40\% of the data as a training set, the next 10\% of the data for the naive model, and the next 10\% as the test set. The third dataset distribution uses the first 60\% of the data as a training set, the next 10\% of the data for the naive model, and the next 10\% as the test set, and the fourth dataset distribution uses the first 80\% of the data as a training set, the next 10\% of the data for the naive model, and the final 10\% as the test set

\section{Results and Evaluation}
We present here the final performance of our models, trained across multiple timescales. In order to obtain upper and lower bounds for each prediction as well as a point estimate, we train three separate models for each timescale, using quantile regression: one to predict the upper bound, one to predict the lower bound, and one to predict the median. We evaluate these models using both standard metrics as well as the custom metrics (overestimation, failed activation rate and  missed activation rate) previously defined in Section ~\ref{sec:eval}.
\begin{figure*}[ht!]
    \begin{subfigure}[b]{0.5\textwidth}
    \centering
    \includegraphics[width=\textwidth]{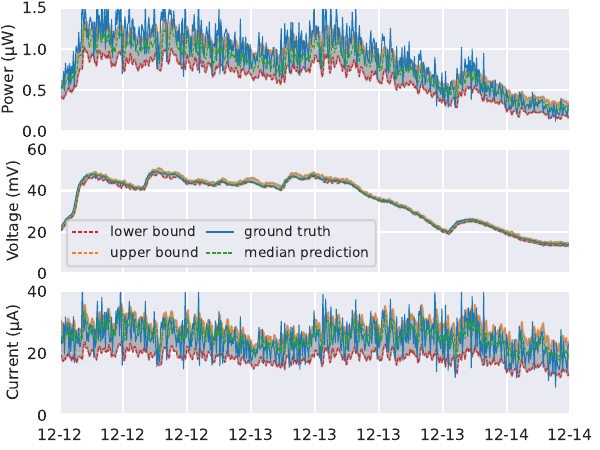}
    \vspace{-1em}
    \caption{5 minute time horizon}
    \label{fig:5_min}
    \end{subfigure}%
    \hfill
    \begin{subfigure}[b]{0.5\textwidth}
  \centering
    \includegraphics[width=\textwidth]{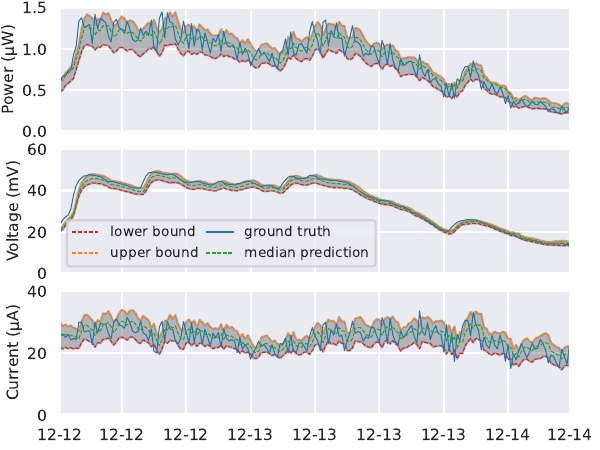}
        \vspace{-1em}
    \caption{15 minute time horizon}
    \label{fig:15_min}
    \end{subfigure}

    \begin{subfigure}[b]{0.5\textwidth}
    \centering
    \includegraphics[width=\textwidth]{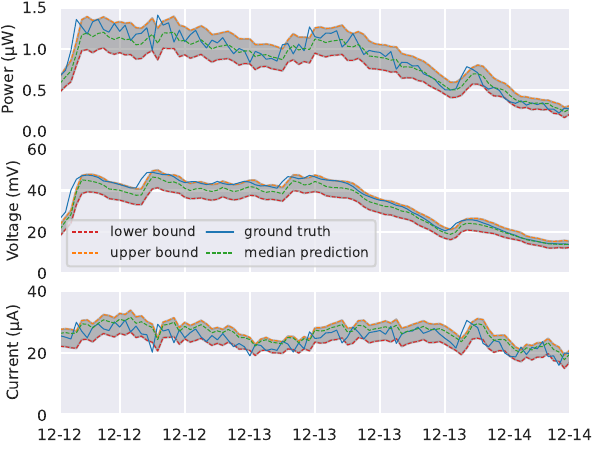}
        \vspace{-1em}
    \caption{30 minute time horizon}
    \label{fig:30_min}
    \end{subfigure}%
    \hfill
    \begin{subfigure}[b]{0.5\textwidth}
    \centering
    \includegraphics[width=\textwidth]{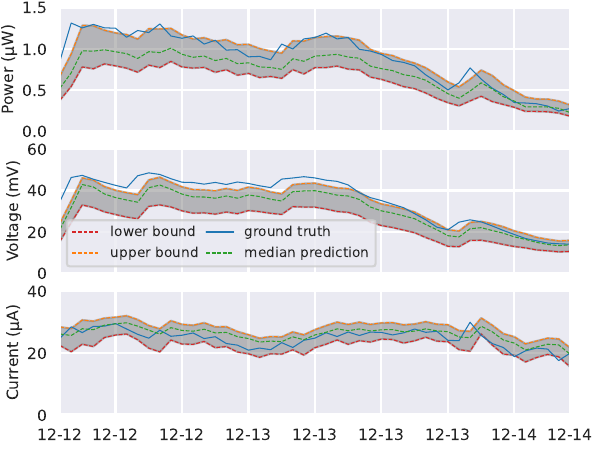}
        \vspace{-1em}
    \caption{1 hour time horizon}
    \label{fig:60_min}
\end{subfigure}

\vspace{3em}
\caption{Estimates and prediction interval plots for various time horizons. Lower and upper bound lines refer to the 5th and 95th percentile predictions, respectively. Plots for 3 minute time horizon omitted, but performance results are available in Table~\ref{tab:model_perf}. These figures present a subset of the data in order to provide a more detailed view of model predictions. Dataset and code used for plotting will be open-source and available on Github and Google Colab.}
\label{fig:all}
\end{figure*}

\subsection{Predictions Across Different Time Horizons} \label{'eval_mod_perf'}
Our models predict average energy generation over 3 minutes, 5 minutes, 15 minutes, 30 minutes, and 60 minutes. In general, models trained at lower timescales result in less wasted energy. Furthermore, the lower-bound models for each timescale allow for the most efficient use of energy compared to the other models of that timescale. For example, when scheduling device activation using our best performing models---the lower-bound 3 and 5 minute models trained on the dataset as described in section ~\ref{dataset}---we only activate the device 0.27\% fewer times than if we had made perfect use of available energy. The lower-bound 60 minute model, in contrast, misses 0.90\% of potential activations. 

It is worth noting that in practice, using a model which makes predictions across smaller timescales would require frequent use of our prediction and scheduling framework, resulting in increased energy expenditure. Even checking the amount of energy currently present in the MFC cell, a neccesary function for this framework, requires energy expenditure. Therefore, a scheduling framework that operates every 3 minutes would have 20 times the operating costs as a model that operates every 60 minutes. Given the low energy generation of SMFCs, increasing energy efficiency from 99.10\% to 99.73\% may not be worth this increased energy cost. It may even be worthwhile to explore the performance of models which make predictions at timescales greater than 60 minutes. Future work will attempt to quantify the costs and benefits across different prediction timescales in more detail, as well as exploring the performance of models which make predictions across longer timescales.

\subsection{Evaluating Model Performance}
Initial results for our prediction and scheduling algorithm for intermittent SMFC-powered device use are promising. Predictions for a subset of our test set are graphed in Figure~\ref{fig:all}. We present and discuss our initial results in this section, and Table \ref{tab:model_perf} contains a complete summary of the performance of each model. 

Using the lower bound of the model which predicts average energy generation an hour into the future, we schedule device activations for 549 hours, or about 23 days. Compared to a naive model which schedules device activations based on the average voltage generated in the past, this framework allows the device to successfully activate a median of 2.08 more times---this is more than a 100\% increase in successful operations. Furthermore, when compared to a theoretical model described in section ~\ref{comp_models} that can perfectly predict and make use of available energy, this framework results in only 4.23\% fewer device activations, if we use the worst-performing lower bound model trained with time-series cross validation. In comparison, when we exclude the models trained on only the first 20\% of the data, the worst-performing lower bound model schedules only 1.13\% less activations than the theoretical maximum

The performance of the scheduling framework generally---but not always---becomes even stronger the lower the timescale of prediction. However, as discussed in Section ~\ref{'eval_mod_perf'}, shorter prediction timescales require increased energy expenditure, and it may not be worth increasing energy cost for improvements in an already high operating efficiency. The tradeoffs between energy consumption and operating efficiency will be further explored in future work.

Another topic for future work will be scheduling different types of device operations, with different functions, energy costs, and consequences for failure. For example, a simple read/record operation could potentially be more aggressively scheduled than a wireless transmission operation, because if the less energy intensive read/record operation fails, the amount of wasted energy is less if the wireless transmission operation had failed. 

\subsection{Time-Series Cross Validation}
In order to test how well our models generalize to different distributions of data, we perform time series cross validation to train and evaluate four different models for each timescale. The distribution of each training, validation, and test set is described in section ~\ref{tscv}. On the whole, the models trained on these data distributions perform fairly strongly. Even the worst performing of the lower-bound estimate models, which was trained on 20\% of the data and made its predictions at intervals of 60 minutes, was able to successfully schedule only 4.23\% fewer device activation than the oracle model defined in section ~\ref{comp_models}. 

\section{Discussion}
\subsection{Importance of SMFC Energy Prediction}
The goal of a predictive model for SMFCs is to allow a system to plan future activities to maximize utilization of the harvested energy. This means wasting as little energy as possible while performing the maximum possible number of useful operations. SMFCs do not produce a great deal of energy, so it is vital that the energy they produce is used effectively. Therefore, it is extremely important for an SMFC powered device to activate only when there is sufficient energy available. If our model over-predicts the energy available at a given time, attempting to activate a device when there is not sufficient energy available, this wastes our carefully stored energy.

To address this need, our model goes beyond the single-point estimations most often used in deep learning, and instead generates a range of feasible predictions for energy production. By considering the confidence intervals on these predictions, we can minimize the possibility of wasting energy by activating a device only when there is a high probability of success.

\subsection{Notes on Performance}
When using the predictions of the lower bound model, our scheduling framework rarely over-predicts how much energy will be available, resulting in a low rate of failed activations for each timescale, as shown in Table ~\ref{tab:model_perf}. It is notable, however, that the lower bound model at the 30 minute timescale has 0 failed activations, despite this metric generally trending upward as prediction timescales increase. The most likely cause of this is that our models predict three values simultaneously - voltage, current, and power - and so they optimize to predict the desired quantile, on average, across these three values. Because of this, for some models, certain predictions will be lower than the desired quantile. It is not completely unexpected for our framework to predict energy generation more conservatively for some models than others, resulting in device activations being scheduled only when there is enough available energy, resulting in no failed activations.

It is also noteworthy that our models are able to predict voltage more accurately than current, particularly at lower timescales. This is likely due to the fact that current typically changes far more from moment to moment than voltage does in our dataset, though it tends to be more stable when averaged across larger timescales. This would also explain why current predictions are far less accurate at smaller timescales than larger ones. Furthermore, since power is the product of voltage and current, it makes sense that power is also predicted less accurately than voltage.

Finally, it is worth noting in the graph of the 1-hour time horizon models in Figure ~\ref{fig:all}, the ground truth voltage and power are greater than the upper bound estimate for the much of the graph. In order to make our graphs more legible, we chose a small subset of the test set to include in the graphs, which happens to include a disproportionate amount of data where the upper bound estimate is lower than the ground truth data for this particular model. However, the upper bound estimate performs much better for the rest of the dataset.

\subsection{Limitations of Current Model}
It is currently extremely difficult to collect reliable, timestamped data on both the power generation of a soil microbial fuel cell and the immediate soil conditions (temperature, volumetric water content, and electrical conductivity). Because of this difficulty, our model is both trained and validated on data gathered from a single microbial fuel cell. As such, it is currently unknown how well this model generalizes to microbial fuel cells deployed in different conditions than the SMFC used for training. Madden et al. \cite{10.1145/3560905.3568110} 
have recently developed specialized logging hardware to gather real-time, accurate data for deployed SMFCs, and this work provides a promising path for making collection of this data easier and more affordable.

\subsection{Future Work}
The operation of the model itself consumes valuable harvested energy. Therefore future work will need to consider the tradeoffs of predicting energy availability and scheduling tasks across various timeframes, as well as accounting for the different types of device operations that would be performed by an SMFC-powered system. The framework scheduling presented by \cite{Zygarde} successfully accounts for a range of tasks for intermittently powered (but not SMFC-powered) systems, outperforming state-of-the-art task schedulers, and we will need to determine to what extent this approach can be adapted to SMFC-powered devices. We will also gather additional SMFC data from cells deployed across a variety of climates, soil types, etc. The broadened dataset will be used in both the training and validation of future models, and will help ensure more generalizability. 

\section{Conclusion}
The prediction and scheduling framework presented in this paper opens the door to a range of future applications for low-power devices fueled by SMFCs. For example, low-power sensors have applications in agriculture, allowing farmers an affordable method to monitor their fields and receive accurate, real-time data on soil conditions \cite{10.1145/3568113.3568117}. Other applications include long-lasting environmental sensors networks designed to detect and prevent environmental disasters, like monitoring conditions across a wide area to help prevent wildfires. In closing, we believe that the work presented here is a valuable step toward increasing the utility of SMFCs, and realizing the goal of an affordable, long-lasting, renewable source of energy.

\bibliographystyle{unsrt}  
\bibliography{main}

\end{document}